\documentclass{aa} 
\usepackage{amsmath,amssymb}
\usepackage[colorlinks = true,
            linkcolor = blue,
                  urlcolor  = blue,
            citecolor = blue,
            anchorcolor = blue]{hyperref}
            
\usepackage{amstext}
\usepackage{natbib}
\usepackage[varg]{txfonts}
\usepackage{graphicx}
\usepackage{arydshln}
\bibpunct{(}{)}{;}{a}{}{,} 

\begin{document}

\title{CTA\,102 in exceptionally high state during 2016--17}

\author
{
Navpreet Kaur \inst{1,2}  \thanks{Email: navpreet@prl.res.in; navp551@gmail.com}
 \and 
Kiran S. Baliyan \inst{1} \thanks{Email: baliyan@prl.res.in; baliyanprl@gmail.com}
}

\institute
{
Astronomy \& Astrophysics Division, Physical Research Laboratory, Ahmedabad 380009, India. \label{1}
\and
Indian Institute of Technology, Gandhinagar 382355, Gujarat,  India.  \label{2}
}

\date{\today}

\abstract
{Blazars, when in outburst, provide a unique opportunity to study their spectral behaviour, correlated variations at different frequencies, and structure of the jet. Such an  unprecedented flaring activity in FSRQ CTA\,102, during 2016 November - 2017 January, is used for a detailed study to understand flaring mechanisms at short and long time scales, spectral behaviour in different energy regimes and to estimate sizes and location of the high-energy emitting region in the jet.
Multiwavelength (MW) data ($\gamma-$ray, X-ray, UV, optical and radio) for CTA\,102 during its outburst  period,  are obtained from Fermi-LAT, Swift-XRT/UVOT, Steward Observatory, Mt Abu Infrared Observatory and OVRO.
 These are analyzed to construct MW light curves, extract the spectral information,  and to perform the correlated variability studies.
Our study shows that CTA\,102 attained the highest ever flux levels across the electromagnetic spectrum (EMS) while flaring and otherwise, with rapid and prolonged activities at all the frequencies. A number of short term (3 to 8 days) and long term (\textgreater month) variability events are noticed across the EMS. We infer a redder when brighter trend in faint state and a bluer when brighter trend during a few optical flares.  
Based on the flux doubling timescale, the size of the $\gamma$-ray emitting region is estimated as $\approx 8.76\times10^{15}$cm, located at a distance of about $5.58\times 10^{16}$ cm from the central engine.
CTA\,102 was in extremely bright phase during 2016-17, due possibly to, successive high energy particle injections into the jet, creating  shocks traveling down the jet which lead to the overall flux enhancement across the EMS.
 Alternatively, a decreasing viewing angle  could also lead to the  enhanced flux.
The study reveals  correlated variations in all the energy bands, with lags within time bins, indicating to co-spatial origin of the emissions. During the flaring event, a bluer-when-brighter color in optical and  harder when brighter trend in X-ray and the $\gamma-$ray spectra are noticed. During some flares  softer $\gamma-$ray spectra are noticed.

}

\titlerunning{CTA\,102: 2016-2017 Outburst}
\maketitle

\section{Introduction}
\label{sec_intro}

Flat spectrum radio quasar (FSRQ), CTA\,102 (also called PKS J2232+114, 4C11.69), at the redshift 1.037 \citep{cta102_z} has been observed across the electromagnetic spectrum (EMS) extensively \citep[][ and references there-in]{Osterman2009}. It was  detected in $\gamma-$rays first time by the EGRET instrument on-board the {\it Compton Gamma-Ray Observatory}  \citep{nolan1993}.  \citet{Tornikoski1999} reported a simultaneous flare in optical and radio (90 GHz) in CTA\, 102 during 1997, while only the optical flare occurred in 1996. The detection of optical variability and the optical-IR correlated emissions, during the high-flux state of the source, were reported by \citet{Raiteri1998} and \citet{Bach2007}, respectively. \citet{Osterman2009} detected the optical micro-variability (aabout 15 min time scale), with as high as 2 mag/day rate of change in brightness and a redder when brighter (RWB) spectral behaviour during the 2004-05 activity in the source. They also found that the variability in radio, optical and X-ray was not correlated, which was in line with the findings by \citet{Osterman2008}  for  the FSRQ PKS 1622-297 and \citet{Clements1995} for a number of FSRQs.   These results indicate that  different emissions  in the jet dominate at different epochs and locations. However, a strong correlation between $\gamma-$ray and optical flaring events in CTA\,102 was reported by \citet{Cohen2014} with $\gamma-$ray leading the optical by about 11 days, suggestive of the leptonic origin of high energy $\gamma$-rays.  The total flux and spectral study by \citet{Williamson2014} found  CTA\,102 in 12 periods of quiescence, 7 of active and 4 periods of $\gamma$-ray flaring state during 2008--2012. The source witnessed small scale variation in its early (1973 -1987) history while it  was rather faint with an average B-band brightness about 17.7 mag which increased by more than three magnitude  during the 2004 outburst, peaking  with R = 14.5 mag on 2004 October 04 \citep{Osterman2009}.    The WEBT campaign \citep{Larionov2016} during the 2012   huge outburst in CTA\,102 captured the source  in its, at that time,   highest flux levels across the spectrum (e.g., R brighter than 13.5 mag, $F_{\gamma}$ = $8.2\times10^{-6} ph cm^{-2}s^{-1}$). A co-spatial origin of the optical and $\gamma$-ray emissions was inferred as the study revealed strong correlated variation in the two energy regimes \citep{Larionov2016} without any lag.  \citet{Casadio2015} used multiwavelength data, including 43 GHz VLBA images,  and found that the correlated optical and $\gamma-$ray flaring occurred at more than 12 pc away from the central source. 

Recently, CTA\,102 underwent unprecedented activity in all the electromagnetic spectrum (EMS) bands during 2016--17 \citep[][ etc.]{at9130, at9788, at9801, at9840, at9841}, significantly surpassing 2012 flare levels and, therefore, attracted lot of attention. The enhanced activity first started in optical, reaching R=14.33 mag on 2016 June 8 \citep{at9130}, while the IR fluxes  significantly increased showing J$=$9.97 mag \citep{at9801}, which was more than 2 mag brighter than that recorded by \citet{Larionov2016} during 2012 flare. \citet{Bachev2017} addressed intra-night optical variations in blazar CTA\,102 during the two extreme outbursts, in 2012 and 2016, and found change in the brightness  by a few tenths of a magnitude on a time scale of a few hours. The fastest variation was noticed within 30 minutes, resulting in a brightness change by 0.2 mag in R band. The multiband optical lightcurves indicated to a strong correlation with out any significant  time lag.  \citet{Zacharias2017} explained the high energy emission during  prolonged 2016 outburst in CTA\,102 on the basis of a  gas cloud entering and leaving the relativistic  jet, causing ablation of the gas and hence gradual increase in the  emissions across the EMS. They put a  cut off limit on the IC component at $\approx$ 20 GeV.  \citet{Raiteri2017} discussed the radio to optical emissions in CTA\,102 during the 2016-17 outburst and explained the variations by invoking an inhomogeneous, curved jet model leading to changes in the viewing angle and Doppler boosting. They also reported emission at 37 GHz peaking much earlier than the peaks in sub-mm (230 GHz) and optical (R-band), inferring different locations for their emission. Based on the short time binnings, 3 hr and 3 mins,  \citet{Shukla2018} detected rapid variability time scale of 5 minutes during the 2017 April flare in CTA\,102, making it one of the very few blazars to show such extreme variability event in high energy $\gamma-$ rays.

Strong correlation between optical and high energy $\gamma-$rays has been noticed in a number of blazars. Using the high energy $\gamma-$ray data from Fermi and quasi-simultaneous optical data, on more than 40 blazars, \citet{Cohen2014} studied the correlation between two fluxes. They found that, in general, FSRQs show high energy emission leading the lower one with a time lag of 1 - 10 days. The behaviour of intermediate (IBL) and high frequency (HBL) blazars  was not that clear regarding the lags and showed small lags or leads. Based on the strong correlation between low and high energy emission for many sources, one zone leptonic models were suggested for the emission. CTA\,102 has shown complex  nature of correlated variations in radio, optical and x-rays during its 2004-05 \citep{Osterman2009} and 2012 \citep{Larionov2016, Raiteri2017} high states. While no significant correlation was seen during 2004, strong correlation was noticed in 2012, also, a significant enhancement at 37 GHz met with quiescent behaviour in optical during 2008-09.  Since it is difficult to understand the multi-band emission behaviour of these sources when in low phase, such outbursts provide opportunity to study correlated variations between various EMS regimes and  to study the structure of the relativistic jet. \citep{Raiteri2012, marscher2010}. 

In this communication, we present a multiwavelength investigation of the blazar  CTA\,102 in its brightest phase in radio, optical, UV, X-ray and $\gamma-$ray energy bands, during 2016--2017 and try to understand its variability behaviour. The next section (section 2) reports multiwavelength observations and data analysis, section 3 discusses the results and the last section, i.e., section 4, concludes the present study.

%Shukla2018 

\section{Multiwavelength Observations and Data Analysis}
\label{sec_data}

\subsection{Fermi-LAT Observations}
\label{sec_fermi}
The LAT is a pair-production telescope onboard the Fermi satellite \citep{atwood2009} with a large effective collection area ($\simeq 6500$\,cm$^2$ on axis for 1 GeV photons) and a large field of view (2.4\,sr),  sensitive in the energy range $20$\,MeV to $300$\,GeV.  The energy range covered is approximately from 20 MeV to more than 300 GeV. The field of view of the LAT covers about 20 per cent of the sky, and maps whole sky about every three hours (192 mins to be exact).
The Fermi-LAT data for the duration 2016 November 1 (MJD 57693) to 2017 January 21 (MJD 57774) were analyzed using ScienceTools software package(version v10r0p5). In our case, we have used the most appropriate event class as 128 with event type as 3(for point and mildly extended sources as suggested by Fermi-LAT analysis team), to analyze the data on the medium to longer time scales. Using the {\it gtselect} tool, we have extracted the photon class events (i.e., PASS 8 SOURCE) lying within the region of interest(ROI) of $10\circ$, zenith angle \textless $100\circ$, energy range between 0.1-300 GeV. The {\it gtlike} tool was used to re-construct the source energy spectrum, using maximum likelihood analysis, and the background model was constructed using 3FGL catalog (${\textit (gll\_psc\_v16.fit)}$ ). The source spectrum was generated using a simple power-law model while we used ${\textit gll\_iem\_v06.fits}$ and $iso\_P8R2\_SOURCE\_v6\_v06.txt$ to model the Galactic diffuse emission and the isotropic emission component, respectively. The Fermi-LAT data has been reduced using the publicly available Python package, Enrico \citep{sanchezdiel2013}. 

We have used the temporal binning of one-day and six hours to three hours, during the whole period (MJD 57693 to MJD 57774) and the major gamma-ray flares(on MJD 57751 and MJD 57760, for sub-hour variability) respectively. The temporal analysis with shorter time bins is constrained by the limited number of photons when source is not bright enough to be detected. Fortunately, in case of CTA\,102 under current unprecedented outburst phase, we had best possible photon statistics (measured by  the Test Statistics parameter,  TS= 2($log L_1$--$log L_0$), where $L_1 \& L_0$ are the likelihood of the data given the model with or without the source, respectively), enabling us to use shorter time bins. We generated the light curves ($lcs$) using 3hr (TS \textgreater 40) and  6hr (TS \textgreater 135) during the two major flares, and 1 day binning (1650 \textless TS \textless 10.5), with more than $3\sigma$ confidence level,  for the whole duration of observations used here, including quiescent phase as well. 

%In all the bins applied, we  have maintained  $\textgreater 3\,  \sigma$ confidence level.

%Nav2017_1es
\subsection{Swift X-ray, UV \& optical Observations}
We used $Swift$-XRT publicly available data from HEASARC database\footnote{http://heasarc.gsfc.nasa.gov/cgi-bin/W3Browse/swift.pl} in the energy range 2  -- 10 keV for the period  from 2016 November 14 to 2017 January 18. The data were processed using HEASOFT package version 6.20.The standard $xrtpipeline$ v.0.13.0  was used to calibrate the data following the defined analysis steps\footnote{http://www.swift.ac.uk/analysis/xrt/}. The source and background spectra were extracted using $xselect$ tool and the pile-up corrections were applied in a few cases when the source was extremely bright. The obtained spectra is fitted with a simple power-law model in XSPEC and the fluxes were calculated, using column density, $ n_{H}= 5.0 \times 10^{20} cm^{-2}$  \citep{dicklock1990}. 
It is to be noted that \citet{jorstad2004} analyzed the Chandra data using $ n_{H} = 3.62 \times 10^{20} cm^{-2}$,  while \citet{stroh2013} found a value of $ n_{H} = (7 \pm 2) \times 10^{20} cm^{-2}$ when analyzing XRT data with an absorbed power law with freely varying $ n_{H}$.

The Swift satellite is a multiwavelength observatory that provides the simultaneous data in optical (U: 345 nm; B:439 nm; V: 544 nm) and ultra-violet bands (UVW2: 188 nm; UVM2: 217nm; UVW1: 251 nm). We used the Swift-UltraViolet Optical Telescope (UVOT; \citet{Roming2005_UVOT}) data for  a period of 2016 November 14 to 2017 January 18, which were reduced using HEASOFT package. Multiple exposures in the same filter at same epoch were summed with $uvotimsum$ and aperture photometry was done with the $uvotsource$ task. The details of the analysis procedure are given by \citet{nav1es2017} and \citet {chandra2015}. 

%{\bf Are the values of of optical/UV corrected for dust absorption (reddening in this direction E(B-V)= 0.0612)}

 The values of  $ \Gamma_{x}$, X-ray photon index,  range between 0.8 and 1.8, with an average value of 1.6 and a harder-when-brighter trend is seen, which is very common in FSRQs \citep[e.g.][] {vercellone2010}. The smaller dispersion of the data points corresponding to the MJD  57740 - 57770 period is  due to their higher number of photons and therefore better photon statistics. 

\subsection{Optical Observations from MIRO}
We monitored the source CTA\,102 on a total of eight nights in December 2016, beginning on December 18, when it was undergoing  unprecedented outburst  across all energy regimes. The optical broadband filters (B ,V, R \& I) were used to carry out high temporal resolution (30--50 seconds per exposure) observations from the  Mount Abu InfraRed Observatory (MIRO) telescope facility-- 1.2 m telescope equipped with ANDOR CCD camera with $2048 \times 2048$ pixels (for details about the instruments and detector used, see, \citet{nav1es2017}). The standard procedures were followed to reduce and analyze the data for the source and available field stars. Their instrumental magnitudes were calculated to generate differential light curves for the source -- comparison star, and, control -- comparison stars in order to determine the existence of any short-term (intra-day) variability.

\subsection{Steward Optical Observatory data}
We also used the optical V-band data and polarization data (DP: Degree of polarization and PA: Position angle) from the public archive at Steward Observatory, Arizona \citep{Smith2009} \footnote{http://james.as.arizona.edu/~psmith/Fermi/}, available only  for the duration from 2016 November 3 to 2017 January 13 for CTA\,102. 

\subsection{Radio data at 15 GHz from OVRO}
The radio fluxes for CTA 102, at 15 GHz, from OVRO\footnote{http://www.astro.caltech.edu/ovroblazars/index.php?page=home} were obtained during the period  MJD 57695 -- 57768.  The observatory, using a 40 m single dish centered at 15 GHz, regularly monitors a sample of 1700 gamma-ray bright blazars, including the sources from CGRaBS (Candidate Gamma-Ray Blazar Survey) and Fermi-AGN catalogs. Each source is monitored twice a week with the typical flux error of 4 mJy and $\approx$ 3\% uncertainty from pointing errors and other systematic effects \citep{richard2011}.

\section{Results \& Discussion}
\label{sec_result}
\begin{figure}
\includegraphics[width=8.7cm,height=9.5cm]{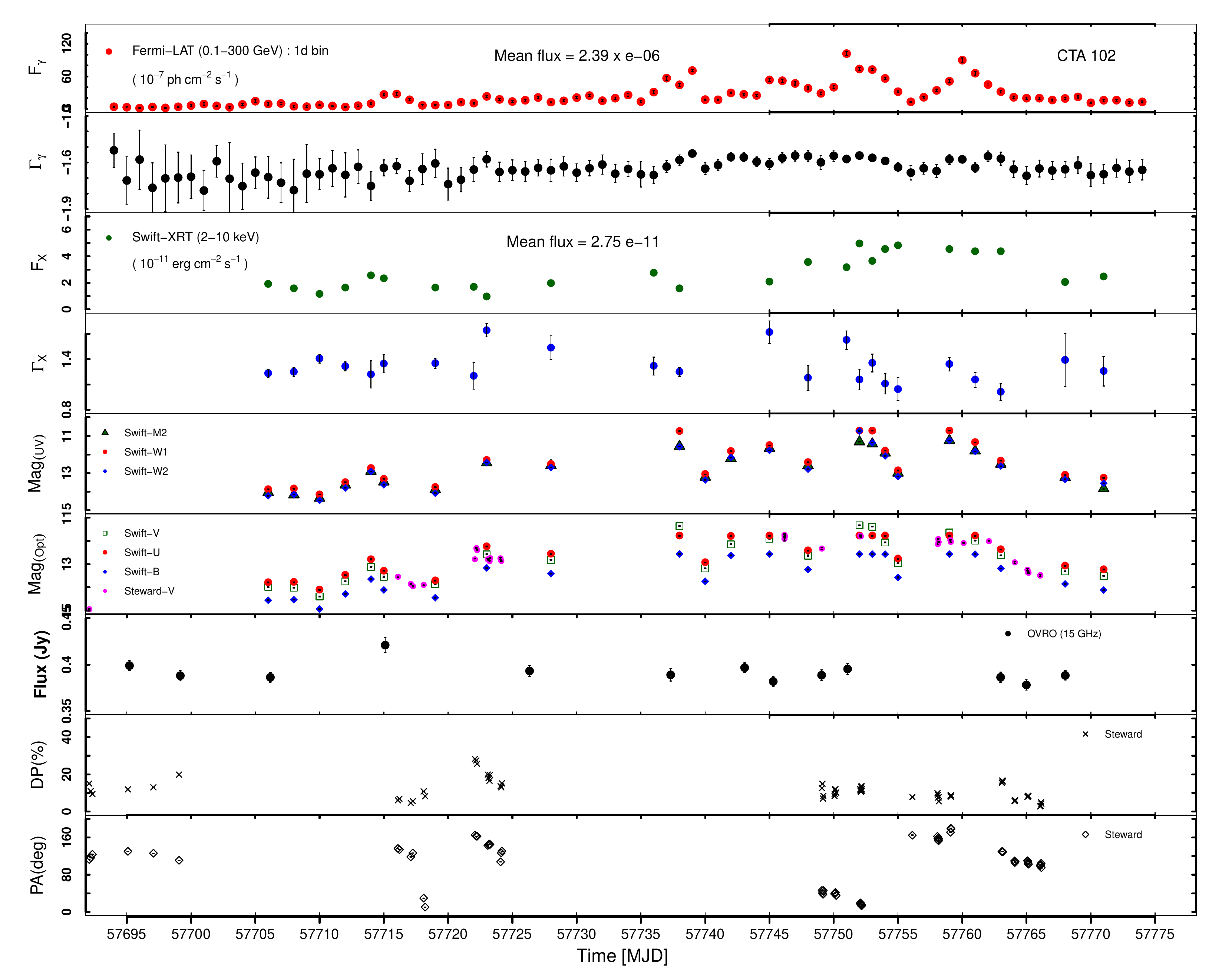}
\caption
{
Multi-wavelength light curves for  CTA\,102 during the 2016 Nov - 2017 Jan outburst.
From top to bottom: 1d binned Fermi-LAT $\gamma$-ray flux (E \textgreater 100 MeV) followed by $\gamma-$ray photon index; 
X-ray  (2.0-10.0 keV) flux (Swift-XRT) ;  X-ray photon index ($\Gamma$; Swift-XRT);
 UV-band (W1, M2, W2) magnitudes (Swift-UVOT);
 Swift-UVOT, MIRO and Steward Observatory optical magnitudes; 15GHz OVRO flux;
Degree of Polarization (DP$\%$) and position angle, PA, in degrees (Steward Observatory).
\label{mwlc}
}
\end{figure}

Figure \ref{mwlc} presents the multiwavelength light curve (MWLC) for CTA\,102 during 2016 November 1 - 2017 January 21. The time in MJD is plotted along X-axis while the respective brightness fluxes/magnitudes are along Y-axis. 
The top panel in Figure \ref{mwlc} shows the $\gamma-$ray flux (in $ ph \ cm^{-2}s^{-1}$), using one day (1d) binning, with an average flux of 2.39$\times 10^{-6} ph\ cm^{-2}s^{-1}$ over the period of the outburst. %which is 300 times larger than that listed in 2FGL catalog. 
Second panel shows $\gamma-$ray photon index, ${\Gamma_{\gamma}}$. In the third and fourth panels we  show the X-ray (2.0-10.0 keV) Swift-XRT flux (in $ergs \ cm^{-2}s^{-1}$), varying between (4.81- 0.96) $\times10^{-11}$ $ergs \ cm^{-2}s^{-1}$ with average flux of 2.75$ \times10^{-11}$ $ergs \ cm^{-2}s^{-1}$ and X-ray photon index.  The Swift-UVOT fluxes in all UV bands i.e., W1, M2, W2, are shown in the fifth panel while sixth panel gives the optical UBV-band magnitudes obtained from Swift-UVOT, MIRO along with Steward Observatory V-band data. Also shown is 15 GHz radio flux (in Jy) obtained from Oven's Valley Radio Astronomy Observatory (OVRO). The bottom two panels display the polarization data (DP and PA) taken from the Steward Observatory. Some of the features are common in all the lightcurves but for more clarity a correlation study is performed between fluxes in  various energy regimes  using statistical technique zDCF described by \citet{Alexander2014}.  The correlation plots are discussed in a later section.

\begin{figure}
\centering
\includegraphics[width=8.5cm,height=5.5cm]{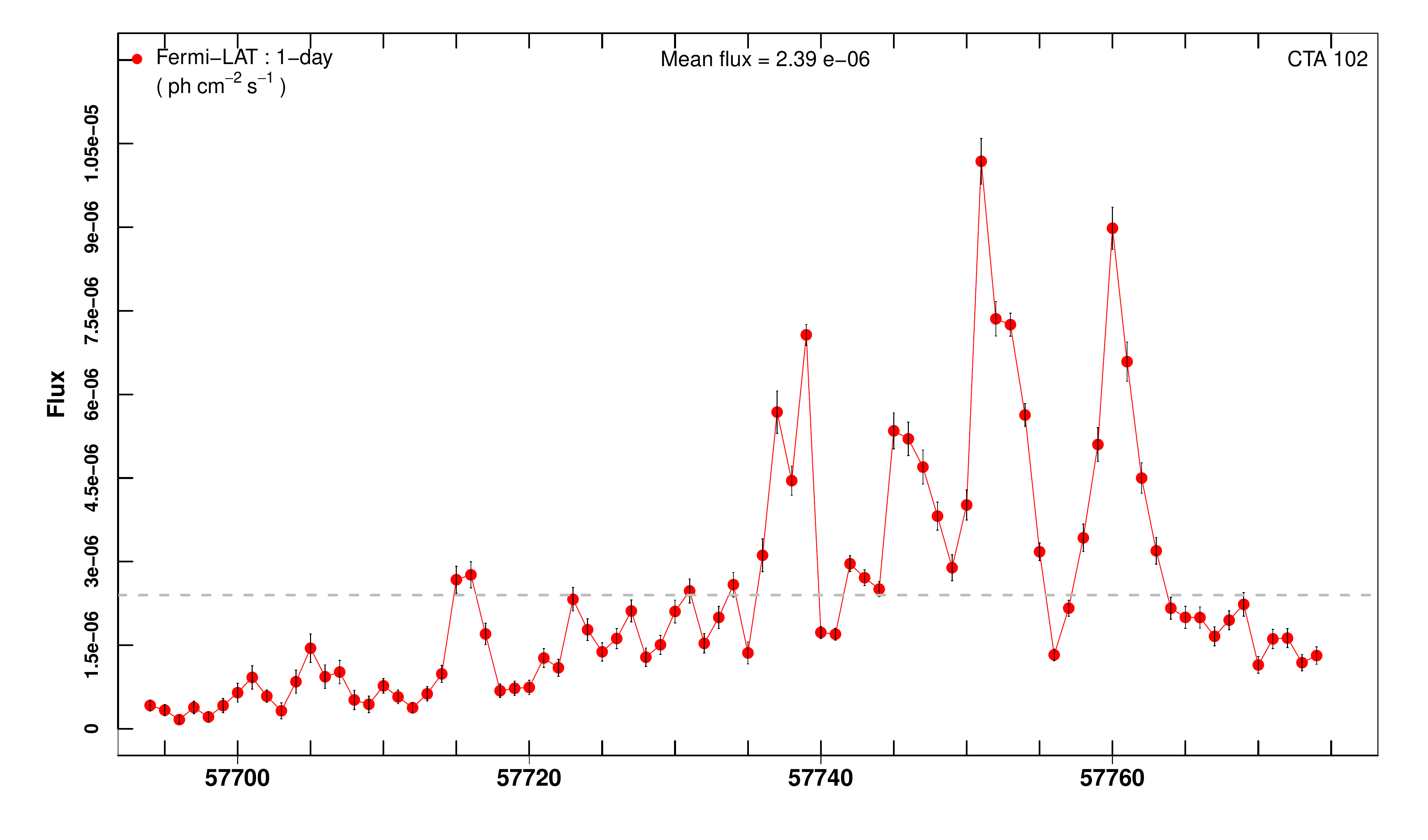}
\caption{$\gamma-$ray light curve of FSRQ CTA\,102 showing flaring activity during November 2016--January 2017. Time in MJD is plotted along X-axis and flux along Y-axis. The grey dashed line shows the average flux level.}\label{gLC}
\end{figure}

The  $\gamma-$ray light-curve shows very interesting features, particularly during outbursts with multiple prominent flares superposed on these outbursts. Therefore, for more clarity we plot $\gamma-$ray(Fermi--LAT) lightcurve in a separate figure (Figure \ref{gLC}), with time in MJD and 1d binned flux in   $ ph \ cm^{-2}\ s^{-1}$ units. During this period (2016 November 1 - 2017 January 21), the average $\gamma-$ray flux (2.39$\times 10^{-6} \ ph\ cm^{-2}\ s^{-1}$)  is more than 300 times of that listed in 2FGL catalog, which is about 2.9$\pm0.2 \times 10^{-9} \ ph\ cm^{-2}\ s^{-1}$\citep{nolan_2FGL12}. To estimate the  duration of outburst, we have to first define the outburst. We call the source to be in outburst phase  if the $\gamma-$ray flux calculated with 1 day binning within the $\gamma-$ray energy range 0.1--300 Gev is larger than  2.0$ \times 10^{-6} \ ph\,cm^{-2}\ s^{-1}$.  It leads to the detection of a big flare at about MJD 57714, lasting just three days with a peak flux of 2.5$\times 10^{-6} \ ph\,cm^{-2}\ s^{-1}$ and the main outburst which started at MJD 57735, with a peak flux of 6$\times 10^{-6} \ ph\,cm^{-2}\ s^{-1}$, which continued for about next thirty days. The strongest flare during the outburst occurred on MJD= 57751, with a peak flux  of 1.1$\times 10^{-5} \ ph\,cm^{-2}\ s^{-1}$. Such strong and prolong  outburst in this source is unprecedented and contributes enormous power to the overall energy flux of CTA\,102.
There are a number of rapid but strong flares superimposed on the already much enhanced base level flux.   
Interestingly, CTA\,102 was so active that during the period considered here, the total flux of the source surpassed  the already high average $\gamma-$ray flux level ($F_{\gamma, avg} =2.39 \times 10^{-6} \ ph \ cm^{-2} \ s^{-1}$) at least nine times. In order to determine extent of the power content in the prominent flares, we calculated the FWHM of each flare profile and multiplied it by the duration of the flares. We found the power contained in the flare peaking at MJD 57739 to be approximately $2.25 \times 10^{-5} \ ph \ cm^{-2}$; for the flare at MJD 57745 about $1.40 \times 10^{-5} \ ph \ cm^{-2}$. The  power contained in the major flare at MJD 57751, is $3.0 \times 10^{-5} \ ph \ cm^{-2}$ while for another major flare at MJD 57760, it is $2.25 \times 10^{-5} \ ph \ cm^{-2}$. The source has not experienced such strong flare with so much power in its entire history.

\subsection{Multiwavelength light curves}
\label{sec_mwlc}
As clearly seen in Figure \ref{mwlc}, the average fluxes of $\gamma-$rays and X-rays are already substantially  elevated and any activity over and  above these levels represent a significant enhancement on their past historical flux values.  It is clearly  noted that CTA\,102 remained very active across all the energies considered here and exhibited a zoo of almost symmetric $\gamma-$ray flares, spanning from 3d to 8d, with significantly high flux levels. At lower energies (UV and optical), an outburst (\textgreater 1 month), superposed by several flares of a few days duration, is noticed. 

CTA\,102 experienced a huge outburst of 70d duration, beginning MJD 57694, across EMS after being in the quiescent state for a long time. The activity started with a short $\gamma-$ray flare (3d long) on MJD 57701, followed by the next flare on 2016 November 13 (MJD 57705) showing a five fold enhancement in the  flux in just 2d ($F_{\gamma}^{peak}=(1.44\pm 0.25) \times 10^{-6} ph\ cm^{-2}s^{-1}$).

The $\gamma-$ray flare with double peaks (1d apart) peaking at MJD 57715/16 (2016 November 23/24) surpassed the average flux level. The corresponding activity at lower-energies, i.e., in optical, UV and X-ray, also started on MJD 57712, peaking one day early (on MJD 57714), with a flux drop when $\gamma-$ray flux was at the peak (on MJD 57716). The radio 15GHz flux peaked couple of days before that in $\gamma-$rays. Though the decay was not captured by Swift in X-ray, UV and V band, MIRO and Steward observations suggest a smooth decay making it a symmetric flare. During the decay of multi-emission flare, linear polarization increases from 5 - 10\% while position angle of polarization undergoes a rotation by 145$^{\circ}$. A blob passing through stationary core in jet could lead to such behaviour. Interestingly, the radio maps from the VLBI-GLAST campaign show the ejection features from the stationary core, during November 28, 2016. This strongly indicates towards the clear injection of the blob into the jet leading to the flaring activity across EMS.
%{\bf check if some features are seen in 37 and 230 GHz- Raiteri work}

Around MJD 57723 (2016 December 1), the overall base level flux kept increasing slowly. While the X-ray flux was low, optical/UV fluxes got enhanced and rapid polarization changes took place (DP: $28\%$ to $13\%$; PA:$165^{\circ}$ to $127^{\circ}$). The changes in DP and PA during $\gamma-$ray flaring have been  studied by \citet{marscher2014} extensively, who explained these to be associated with a blob passing through a quasi-stationary core, resulting in the significant emissions at all frequencies under shock-in-jet scenario. The blob/shock compresses the plasma in the dense medium of the core, aligning the magnetic field which results in change in polarization and position angle.  We noticed a significant delay between optical/UV and $\gamma-$ray fluxes.  The low X-ray flux could either be the victim of differential alignment of respective emission regions w.r.t. LOS \citep{raiteri2011}, or its origin itself could be different \citep{Cavaliere2017}. Due to large data gaps in X-rays on these epochs, it is difficult to draw any definite conclusion.

A host of rapid $\gamma-$ray flares, with the "saw-tooth" features (slow rise-fast decay), are seen with peaks close to the average flux level on MJD 57727, 57731 and 57734 (2016 December 5, 9 \& 12).

Figure \ref{mwlc} clearly shows the twin $\gamma-$ray flares on MJD 57737 \& 57739, with the flux levels of $(5.68\pm0.37) \& (7.06\pm0.18) \times 10^{-6} ph\ cm^{-2}s^{-1}$, respectively. The X-ray flux increased by  more than 50\% relative to the flux on MJD 57736, followed by a $\gamma-$ray flare (at MJD 57337).  Before the second $\gamma-$ray peak (at MJD 57739), the X-ray flux dropped by  about 44$\%$, followed by the decay in UV \& optical. On MJD 57739, when the fluxes at lower energies were slowly declining,  a $\gamma-$ray flare (second peak) indicates a possible orphan flare. The cause behind such flare could be up-scattering of the ambient photons at the boundary of the jet without any change in the optical flux . There are other explanations based on hadronic scenario for the high energy emission production\citep{Bottcher2006}.
\citet{MacDonald2017} explain orphan flare based on their, ``Ring of Fire'', model in which synchrotron electrons in the jet spine blob up-scatter seed photons emanating off a rind of shocked sheath plasma enshrouding the jet spine. As the emitting blob propagates through the ring, scattering of ring photons by blob creates orphan flare.  It was used to explain 2011 June 11 orphan flare in CTA\,102 along with such flares in 3C273, 3C279, 4C 71.07 etc.

During MJD 57740-57749 we detect a triple-flare, successive constituent flares showing increased amplitudes. Out of these, the major flare peaked at MJD 57745 (2016 December 23) 
with a fast rise (1d) and slow decay (4d). This flare is fit by \citet{Shukla2018} to arrive at a 2hr time scale of variation. The X-ray showed subdued activity, peaking 3d later while the optical and UV fluxes were enhanced by 1 mag $\&$ 1.7 mag, respectively. Radio 15GHz flux seems to have peaked two days prior to activity at higher energies. The major flares during MJD 57751 and 57760 are detailed in the following section.

\smallskip

\subsubsection{Historical outburst with sub-day variability}

	As we noticed from the Figure \ref{mwlc}, CTA\,102 exhibited unprecedented flux levels almost across the EMS. It is accompanied by significant variations in flux at various time scales.  For blazars, detection on intra-day variations have been common in the optical and near infra-red. However, the exceptionally high flux levels obtained for CTA\,102 have made this source one of the very few for which it has been possible to address sub-day time scale variations in high energy domain. It is due both to the  small telescope size (due to space payload limitations) and intrinsically low flux of the high energy sources. However, very fast variations have been detected with a few minutes time scale in TeV regime for a number of sources, in general, BL Lacs \citep[][and references there in]{Ahronian2003}. Here we specifically discuss the variability in optical and the high energy  $\gamma-rays$   at intra-night timescales.
% aharonian2003>> ahronian2003 elsewere

\smallskip
\noindent{\bf Optical intra-night variations}

\smallskip
	In the past, CTA\,102 showed rapid variations in optical with time scales as short as 15 min to 3.6 hrs \citep{Osterman2009, Raiteri1998, Bachev2017}. Very recently, \citet{Zacharias2017} also claimed sub-day variability in optical window with a brightest magnitude of 10.96$\pm0.05$ in R band on 2016, December 29 (MJD 57751).  %and the details are shown in the table \ref{table:mirocta102}. 
 The optical observations from MIRO show that the source has undergone a significant flux variability over short timescales (day-to-day) also. We calculated the nightly averaged R--band magnitudes for CTA\,102 from our observations over 8-nights, beginning 2016 December 18,  and the results are provided in the Table \ref{table:mirocta102}, where columns 1 and 2 represent the epoch of observations in dd-mm-yyyy format and in MJD format, respectively. Column 3 \& 4 show the number of images per night and duration of observations in minutes. The nightly averaged R--band magnitudes and their photometric errors are listed in columns 5 and 6, respectively.

\begin{table*}
\centering
\caption{Nightly averaged R-band optical brightness magnitudes for the  blazar CTA\,102, obtained using 1.2 m MIRO telescope, during December 2016. The epoch of observation is in (dd-mm-yyyy) format and  N is number of images. \label{table:mirocta102}}
\begin{tabular}{cccccc}
\hline
\hline
Date		&MJD	& N & Duration & R-band & $\sigma_{R}$\\
(dd-mm-yyyy)    &     &  & (mins)  & (mag)  & (mag) \\
\hline 
18-12-2016 &57740.66 & 105 & 70.00 &12.68 &0.01 \\
19-12-2016 &57741.60 & 63 & 52.50 & 11.55 & 0.01 \\
20-12-2016 &57742.57 & 122 & 81.33 &  11.84 &0.01\\
21-12-2016 &57743.58 & 93 & 62.00 & 11.99 & 0.01 \\
22-12-2016 &57744.57 & 62  & 41.00 & 11.88 & 0.01\\
27-12-2016 &57749.62 & 4  & 3.34 & 11.23 &0.01\\
28-12-2016 &57750.53 & 6  &  5.00 & 10.98 & 0.01\\
29-12-2016 &57751.57 & 72 & 48.00 & 10.92 & 0.01\\
\hline	                 
\end{tabular}
\end{table*}

Interestingly, CTA\,102 brightened by more than a factor of 2.5 within a day, from  R-band magnitude of  R$=$12.67$\pm$0.01 (2016 December 18) to a value R$=$11.55$\pm$0.01 (December 19), which is unprecedented in the history of this source. After that, the source brightness decreased slowly during next three days with respective magnitudes as 11.84$\pm$0.01 (December 20, 2016), 11.99$\pm$0.01 (December 21, 2016), 11.88$\pm$0.01 (December 22, 2016). As per our observations on December 27, 2016, the source entered into the extreme flaring phase with  a 0.6 mag enhancement from the previous day brightness value, with the nightly averaged magnitudes of R$=$11.23$\pm$0.01. The CTA\,102  continued brightening further with R$=$10.98$\pm$0.01 on December 28, 2016. The very next day, 2016 December 29, the source attained the optical magnitude of R$=$10.92$\pm$0.01, which represents the historically brightest level ever achieved by the source. During this period, the source was detected with a significant intra-night variability (INV),  by \citet{Bachev2017}, recently. While 0.4 - 0.5 mag changes were noticed within a few hours, fastest variability in optical was recoded with 0.2 mag change in just 30 minutes.  During high activity in 2004-05, \citet{Osterman2009} detected microvariability which was absent in radio. They claimed that microvariability was not related to the state of the source. 

 The source, therefore, shows strong day-to-day variations, as revealed by our observations and those by others.  Blazars are understood to be heavily jet dominated sources, hence events on the accretion disk should not affect the changes in the flux, particularly when the source is very active. In that case the physical processes responsible for rapid variations of the continuum flux in the inner  jets should be instabilities in the particle acceleration mechanism, variations in the electron injection, small-scale inhomogeneities in the magnetic fields or the jet plasma.

 \citet{Bachev2017} carried out multi-band optical study of 2012 and 2016 flares in CTA\,102. The authors found no variability in Mg II lines over few years duration. Any correlation between the nuclear flux variations and line emission would indicate nuclear emission being reprocessed by BLR. A time delay between the variations in the nuclear flux and flux in the Mg II line would enable estimation of the mass of the central black hole.  The authors report brightest state of the blazar with intra-night variability of 0.2 mag within about 30 minutes. The variations are reported  to be due to change in the Doppler factor of the blobs. Using this variability time scale in optical, we can estimate the size of emission region from, 

 \begin{equation}
  R \textless  \frac{\delta c \tau}{(1+z)}
 \end{equation}

 where c, $\delta$ =35 \citep{Casadio2015} , and z(1.037) are the speed of light, Doppler factor and redshift, respectively. The size of the optical emission region, with variability time scale as 30 minutes,  is estimated to be $9.28\times 10^{14} cm$. Now, since the mass of the black hole for CTA\, 102 is $8.5\times 10^8 M_{\odot}$ \citep{Zamani2014}, the radius of event Horizon will be about  7.65$\times10^{14}\  cm$ which is of the order of the upper limit on the radius of emission region. It is, therefore, more likely that these rapid variations are originated far from SMBH at parsec or larger distances cause by small scale inhomogeneities or interaction of shocks moving down the jet with particle over-densities. Several blazars, including CTA\,102 \citep{Larionov2016}, are known to show strong correlated flares in optical and $\gamma-$rays, generally linked to passing of relativistic blob through the radio core \citep{marscher2010, Agudo2011}.

 \smallskip
 
 \noindent{\bf Sub-day activity in $\gamma-$ray regime}

 \smallskip
The $\gamma-$ray flux from CTA\, 102 also exhibited significant intra-day and day-to-day variations.
The most pronounced 5d $\gamma-$ray flare, with unprecedented flux levels in the history of CTA\,102, occurred around MJD 57751 (2016 December 29). The $\gamma-$ray flux, $F_{\gamma} = 1.02\times 10^{-5} ph\ cm^{-2}s^{-1}$, larger by about 40\% than the one recorded during 2012 flare as reported by \citet{Larionov2016}, increased by \textgreater   2.5 times in two days.  This flare, along with its shoulder flare almost overlapped with it, cooled down during the next 5 days. The counterparts of this flare in the optical, UV and X-ray light curves show a lag by one day (cf. Figure \ref{mwlc}). The X-ray flux doubled in one day causing a sharp flare over already enhanced X-ray flux while optical and UV brightness increased by more than 1.5 \& 2 mags, respectively, making these flares unprecedented across EMS.
\smallskip
	
	An almost symmetric major $\gamma-$ray flare of 8d duration took place, centered around MJD 57760 (2017 January 7), with a peak flux of 8.98$\pm0.37 \times 10^{-6} ph\ cm^{-2}s^{-1}$ (higher than 2012 levels) and $\Gamma = -1.57\pm0.02$ following, with 1 day lag, the optical and UV flares with 1 mag and 2.8 mag enhancements, respectively. The X-ray was still showing  a plateau while UV decayed slowly; the optical flux remained high for another three to four days followed by a rapid, albeit smooth, flux decrease. Not much activity was seen in polarized flux, but the position angle changed by more than 150$^\circ$. After this high activity, $\gamma-$ flux dropped with a minor flare (on MJD 57769; 2017 January 16) towards the end, with a slightly enhanced X-ray flux.

\begin{figure}
\centering
\includegraphics[width=4.3cm,height=4.2cm,]{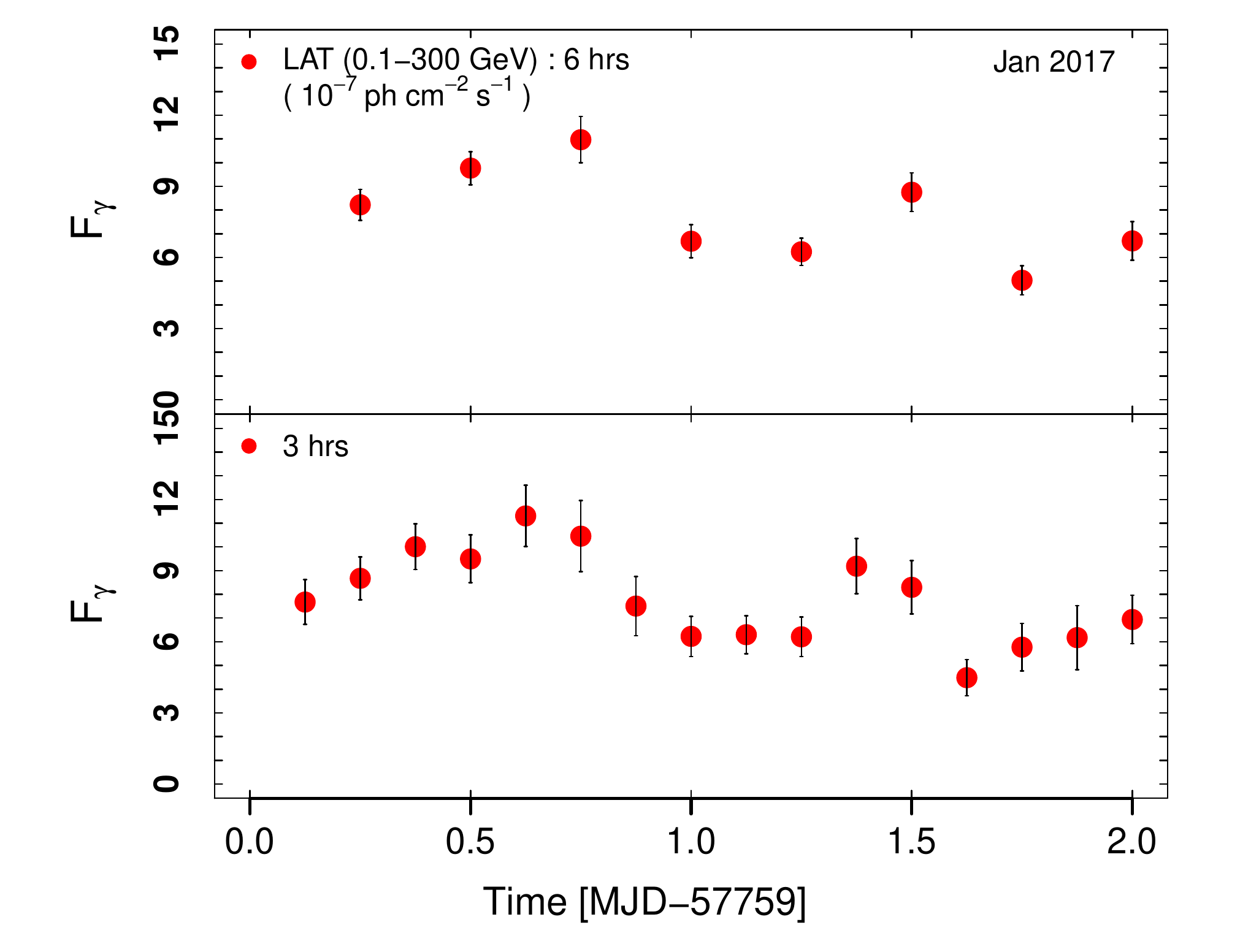}
\includegraphics[width=4.3cm,height=4.2cm]{Fig3b.pdf}

\caption{$\gamma-$ray intra-day flux variability in CTA\,102 using 3 hrs and 6 hrs binned data, for the two major flares on MJD 57750-57752 (left: December 2016) and MJD 57759-57761 (right: January 2017).}\label{inv}
\end{figure}

The extreme flux levels reached during the present outbursts prompted us to use smaller time bins, i.e., 6 hr and 3 hr (TS \textgreater 10) in the data in order to investigate rapid variability at shorter timescales during the two major $\gamma-$ray flares, centered at MJD 57751 (2016 December 29) and MJD 57760 (2017 January 7). Figure \ref{inv} shows the  $\gamma-$ray $lcs$  for the major flares during 2016 December (left) and 2017 January (right), with the upper panel showing 6 hrs and lower one 3 hrs binned data. Clearly, the source showed significant flux variability at sub-day time scale.  The variability amplitudes were estimated using the relation,
	\begin{equation}
	 A_{var} = \sqrt{(A_{max}-A_{min})^2 - 2 \sigma^{2}}
	\end{equation}
	where, $A_{max}$ and $A_{min}$ are the maximum and minimum $\gamma-$ray fluxes respectively, and $\sigma$ is the standard deviation.
 
 The amplitude of variability, $A_{var}$, is estimated to be  6.70 (1.93) $\times10^{-6} ph\ cm^{-2}s^{-1}$ on MJD 57751 and $6.25 (1.82) \times 10^{-6} ph\ cm^{-2}s^{-1}$ on MJD 57760, with \textgreater 3$\sigma$ confidence. The shortest time scales can be used to put upper limit on the size of the $\gamma-$ray emitting region. To estimate the size of the $\gamma-$ray emission region, we used the doubling/halving timescales as calculated from,
 
 \begin{equation}
  F(t) = F(t_0) 2^{\frac{t-t_0}{|\tau|}}
 \end{equation}

 where, F(t) and $F(t_0)$ represent the fluxes at time $t$ and $t_0$ and $|\tau|$ is the halving time scale. The 3-hr binned light curves during 2016 December 29 flare shows rapid variation and is used to get halving time scale. Using the above expression, we have estimated the shortest time scale of variability as $\tau$=4.72 hrs.

 The shortest time scale of variability provides important information about the size of the emitting region. The timescale, calculated above, is used to constrain the $\gamma-$ray emission region size,
  estimated using the same expression (Eq.1) as used for optical emission size above. The emission size is  estimated to be $8.758 \times 10^{15} cm$. Recently, \citet{Shukla2018} used 3 minute data bins from the 2018 April observations of CTA\,102 and obtained a time scale of about 5 mins, which is much shorter than the SMBH horizon light crossing time. The $\gamma-$ray emitting  region, therefore, should be located down the jet far away from central engine \citep{Tavecchio2010}.

 One of the major issues facing the blazar jet study is poorly understood structure of the jet-- the manner in which matter responsible for energy dissipation is distributed along the jet.  The determination of the location of high energy emitting region is possible, when the source is highly variable at those energies (preferably, in flaring state). The flaring or outbursts in blazars are rare phenomenon as most of the time they remain in quiescent or low flux state. Therefore, there are very less number of sources for which the  location of the high energy emission is determined \citep[ and references there-in][]{nav1es2017,larionov2013,wu2018}. The rapid variability in $\gamma-$rays, with a time scale of a few minutes to a few tens of minutes, suggests the emission region to lie close to central engine, within a parsec \citep{Tavecchio2010}.  The problem in this case is that the $\gamma-$ray emission thus produced will get absorbed by the optical/UV photons of BLR region \citep{LiuBai2006, Poutanen2010}, jet or accretion outflow \citep{Dermer-Schlickeiser1993, marscher1992}.  As a solution to this, \citet{marscher2010, Tavecchio2010} proposed that the rapid variability could also be produced in the jet far away from the black hole if the emission region occupies only a small fraction of the jet instead of the whole cross section (jet-in-jet scenario). The strong correlations detected between $\gamma-$ray and mm-wave emissions are suggestive of the emitting regions at more than parsec away from the base of the jet, well beyond BLR. It has been seen in several sources that the $\gamma-$ray outbursts are triggered by the passage of superluminal blobs (knots) through the mm-submm VLBI core. \citet{Casadio2015} reported such trend in the data from 2007- 2014 in CTA\,102 with a suggestion that correlated flare occurred at about 12 pc away from SMBH, but only when motion of emitting regions coincided with the LOS. The issue, however, is still debatable.
 
%The flares in this case were accompanied by increased flux in X-ray and optical energy bands of CTA\,102.

%%% NAVPREET- PUT REFERENCE HERE OF THE PAPER WHICH REPORTS ESTIMATION OF DISTANCES TO EMM REGION BASED ON RAPID VARITION FOR 100s of BLAZARS here...

Recently \citet{yan2017} estimated distances from the black holes to the dissipation regions from where $\gamma-$ray emission originated for the two blazars (PKS 1510-089 and BL Lac), based on the variability time scales. Here, we use doubling time scale as shortest characteristic time scale of variability obtained for the $\gamma-$ emission. To have an estimate of  the distance to the $\gamma-$ray emitting region from the central super-massive blackhole, we need the information about the opening angle close to the base of the jet,  Doppler factor ($\delta$), flux doubling time scale ($\tau_{d}$) and the redshift (z).  The jet opening angle for blazars is generally less than 1 degree, as discussed by  \citet{Jorstad2005}, in general. \citep{Pushkarev_iX_2012} also put 1.0$^\circ$ as the upper limit of opening angle for BL Lacs. We, therefore, estimate the distance to the location of $\gamma-$ray emission from central engine, using following relation,
 
 \begin{equation}
 d= \frac{\delta c \tau}{(1+z) \theta_{j}} 
 \end{equation}

 where, the jet opening angle is, $\theta_{j}\approx  0.7^{\circ}$  \citep{Jorstad2005}. We find that the  $\gamma-$ray emitting region is located at a distance of $d=5.58\times 10^{16}cm$ from SMBH, which is near the boundary of  BLR (R=6.7$\times 10^{17}$ cm) dissipation region \citep{Shukla2018}.

\citet{Fromm2015} estimated distance of the black hole from 86 GHz core in CTA\,102 as about 7 pc using an opening angle of 2.6$^\circ$ and a value of 35 for the Doppler factor, while $\gamma-$ray emission was produced at a distance of 12 pc away from black hole\citep{Casadio2015}. This appears to be more plausible region considering one can avoid pair production on BLR photons.

 \subsubsection{Fractional variability amplitude}
 
 In blazars, the variability is largely stochastic event in nature at all frequencies and timescales, particularly at shorter ones as these are governed by the relativistic jet processes. The similarities and differences among flare profiles reflect varying extents of the particle acceleration and energy dissipation. The amplitudes of variation would depend upon the strength of magnetic field, viewing angle, particle density and the efficiency of acceleration. The availability of good quality data across the EMS makes it possible  to determine the variability amplitude at all the energies. This could be determined using the fractional root mean square (rms) variability parameter, introduced by \citet{Edelson-Malkan1987, Edelson-Krolik-Pike1990}. Basically, the excess variance was used to compare the variability amplitudes at different energy bands from the same observation \citep{vaughan2003}. This methodology has some limitations as it only works  nicely for the densely sampled data having small flux uncertainties at different energies, obtained from different instruments with different but high sensitivities. The large data gaps or high uncertainties in the flux measurements are capable of introducing larger values of rms variability amplitudes rendering the method useless. In the present case, the availability of good quality data across EMS allowed us to determine the variability in the source. The fractional variability and the associated errors are calculated using the relation by \citet{vaughan2003},

\begin{equation}
  F_{var}=\sqrt{\frac{(S^2 - \sigma^2)^2}{r^2}} 
\end{equation}

\begin{equation}
 \sigma F_{var}=\sqrt{(\sqrt{\frac{1}{2N}} \frac{\sigma^2}{r^2 F_{var}})^2+(\sqrt{\frac{\sigma^2}{N}} \frac{1}{r})^2}
\end{equation}

where, $S^2$ is the sample variance, $\sigma^2$ is the mean square value of uncertainties associated with each observation, and $r$ is the sample mean. 

\begin{table}
\centering
\caption{The fractional variability amplitudes (F$_{var}$) for the  blazar CTA\,102 in different energy regimes.\label{cta102_fvar}}
\begin{tabular}{lcc}
\hline
\hline
Energy regime		& $F_{var}$ & $\sigma_{F_{var}}$\\
% (dd-mm-yyyy)    &       & (mag)   \\
\hline 
$\gamma-$ray & 0.873 &0.009 \\
X-ray & 0.459 & 0.001 \\
UVOT- W1 & 0.089 &0.002\\
UVOT- M2 & 0.077 & 0.002 \\
UVOT- W2 & 0.082 & 0.002\\
UVOT-  V  & 0.073 &0.003\\
UVOT- U  & 0.063 & 0.002\\
UVOT- B  & 0.059 & 0.002\\
\hline	                 
\end{tabular}
\end{table}

The calculated values of the fractional variabilities thus obtained are given in Table \ref{cta102_fvar}. We find that the source exhibits larger fractional variability ($F_{var}$) towards higher energies ${\it i.e.}$, 0.87 in $\gamma-$rays, 0.45 in X-rays, 0.82 in UVW2-band, 0.059 in optical B-band. It shows that during the outburst period, jet emission was highly dominant. The large fractional variability towards higher frequencies could be due to large number of particles producing high energy emission. The internal shock model also predicts the high intrinsic amplitude of variability towards higher frequencies.  Particularly low value of the fractional variability in optical could have been affected by the thermal contamination, though in the high flux state, its effect becomes negligible as noticed by \citet{Larionov2016}. On the other hand, \citet{bonning2009} reported $F_{var}$ decreasing towards higher energies for FSRQ 3C454.3. It was due to significant steady thermal emssion (BBB) from accretion disk and line emission in relatively low phase.

\subsubsection{Color and spectral behavior}
It is interesting, and informative, to investigate how the color changes with flux variations in blazars. In general BL Lacs show bluer when brighter (BWB) color in optical supporting shock-in-jet model for the flux variations. On the other hand, FSRQs show redder when brighter (RWB) trend \citep{bonning2009}. For the activity in CTA\,102 during 2004-05, \citet{Osterman2009} found RWB with rapid variations in brightness -- 0.06 mag change within 15 minutes. They also noted BWB behaviour when source was in faint state due to excess in UV, big blue bump (BBB).  It should be understood that BBB and highly beamed synchrotron emission from the jet contribute quite differently during the outburst and quiet phases of the source. In addition, for FSRQs, thermal contribution from the line emission, MgII $\lambda$ 2800 $\AA$ in this case, will also contribute, depending upon the location of emission region and strength of the variable continuum. \citet{Larionov2016} found no change in the strength (effective width) of the Mg II line during the outburst. Thus enhanced activity in the jet had negligible effect on BLR. It could be due to flare happening far from BLR or it was partly due to change in geometry.

In the present study, we plotted (B-V) v/s B, (U-V) v/s U, (W2 - W1) v/s W2 and (M2-W1) v/s M2. We have also plotted X-ray and $\gamma-$ray photon index (see, Figure \ref{mwlc}) along with lightcurves. We studied the spectral behaviour of the source CTA102 with respect to the source brightness and with time, during 2016 November-December - 2017 January in the optical and UV energy regimes. In the color plot between B-V versus B-mag, we noticed the sudden large changes in the B-V color during the flaring state, with 0.5 - 0.9 mag color change when source brightness changed from 12.38 to 12.45 mag. The source showed  a mild BWB in the range 12.45 - 13.7 mag and RWB towards fainter than 14 mag.  The overall trend revealed a mixed behavior of the source in different brightness states. We did not notice any significant change in the color with time.  
We have noticed the similar spectral behavior in U-V versus V-mag.
In the UV-regime (W2-W1 versus W2-mag), we noticed a significant BWB trend during the bright state of the source while a mild RWB trend, when W2\textgreater 12.5 mag, is seen. The (W2-W1) color shows drastic BWB behaviour during flaring. However, no trend is noticed on the average. This could be due to the enhanced high energy particle density due to the passage of  shock down  the jet.
In the present study, we noticed that the source is getting bluer during the flaring epochs and shows mixed  behaviour otherwise. In general we see BWB when source is in high flaring phase while a RWB color when source is relatively faint, while in-between, it either shows RWB or no trend at all.  \citet{Osterman2009} report an RWB trend when CTA\, 102 was in high state while \citet{Bachev2017} \& \citet{Zacharias2017} notice constant colors with time during intra-night variations. \citet{Gu2006} suggest RWB behaviour to be due to the presence of BBB in FSRQs, which could be significant in faint optical state and gets washed out when  Doppler boosted jet emission dominates. Taking into account the spectral properties of flaring blazars, in general, the spectral trend from RWB to BWB is a kind of transition from FSRQ to BL Lac, where the synchrotron emission from the jet is dominant.

In case of $\gamma-$rays, we notice a harder when brighter spectral trend, similar to what \citet{Shukla2018} have noticed. However, during the decay part of flare peaking at MJD= 57745, photon spectral index gets softer (increases) which continues up to the onset of next major flare.  The X-ray photon spectral index anti-correlates with X-ray flux - spectra become harder when source is brighter, albeit photon index shows rather harder state while flux is almost constant during MJD 57758 - 57763. 

The optical spectral analysis showed a mild redder when brighter (RWB) color, in general and a bluer when brighter (BWB) trend during the flaring period. We notice that the spectral variability is more profound at higher energies with X-ray becoming harder while $\gamma-$ray spectra getting softer during the flaring period as compared to the whole period considered here.

 Using the luminosity distance ($d_L $) and the brightest X-ray flux value ($F_X$), we estimated the X-ray jet isotropic luminosity 
to be $L_{X,jet}^{iso}=  2.68 \times10^{47}$ erg/s, which exceeded the Eddington luminosity ($L_{Edd} = 1.11\times10^{47}$ erg/s; \citet{Fromm2015}), with $\eta = L_{X,jet}/L_{disk} =6.4$ in the source during its brightest phase, requiring a larger Lorentz factor.
%If we consider the accretion efficiency $\eta$=0.01 (i.e., 10$\%$) and $L_{X,jet}^{iso}$, we get the accretion disk luminosity as $L_{disk}$=$ 2.68 \times10^{49}$ erg/s, which exceeds the jet luminosity by order of two magnitudes. 

\subsection{Correlated variations  using zDCF}
We have also studied the correlated flux variations  between two $LC$s obtained for different energy regimes. These provide the information on how well the variability in two energy bands matches and allows one to measure the time it takes for one emission region to respond to the changes in the other.  From the Figure \ref{mwlc}, it can be noticed  that, broadly,  all the fluxes in different energy regimes vary in-tandem. The 15 GHz radio flux has been on the decrease since it outburst phase during 2012 activity, apart from occasional enhancements. During MJD 57715 flare, optical/UV/X-ray lead $\gamma-$ ray and radio, $\gamma-$ray, optical/UV peak at MJD 57740 when radio flux is increasing. The flux variations  at 37 GHz and 230GHz follow those in R-band \citep{Raiteri2017}. To investigate  it further quantitatively, we used the discrete correlation function \citep{Edelson1988} to analyze correlated variations among light curves. The whole period of our consideration, which largely covers the outburst period 2016-17,   is dominated by a number of significant flares. These flares are possibly produced by the shocks moving down the inhomogeneous jet or by the change in the viewing angle, caused by helical jet motion, leading to variation in the Doppler factor. 

\begin{figure*}
\centering
\includegraphics[width=0.3\textwidth]{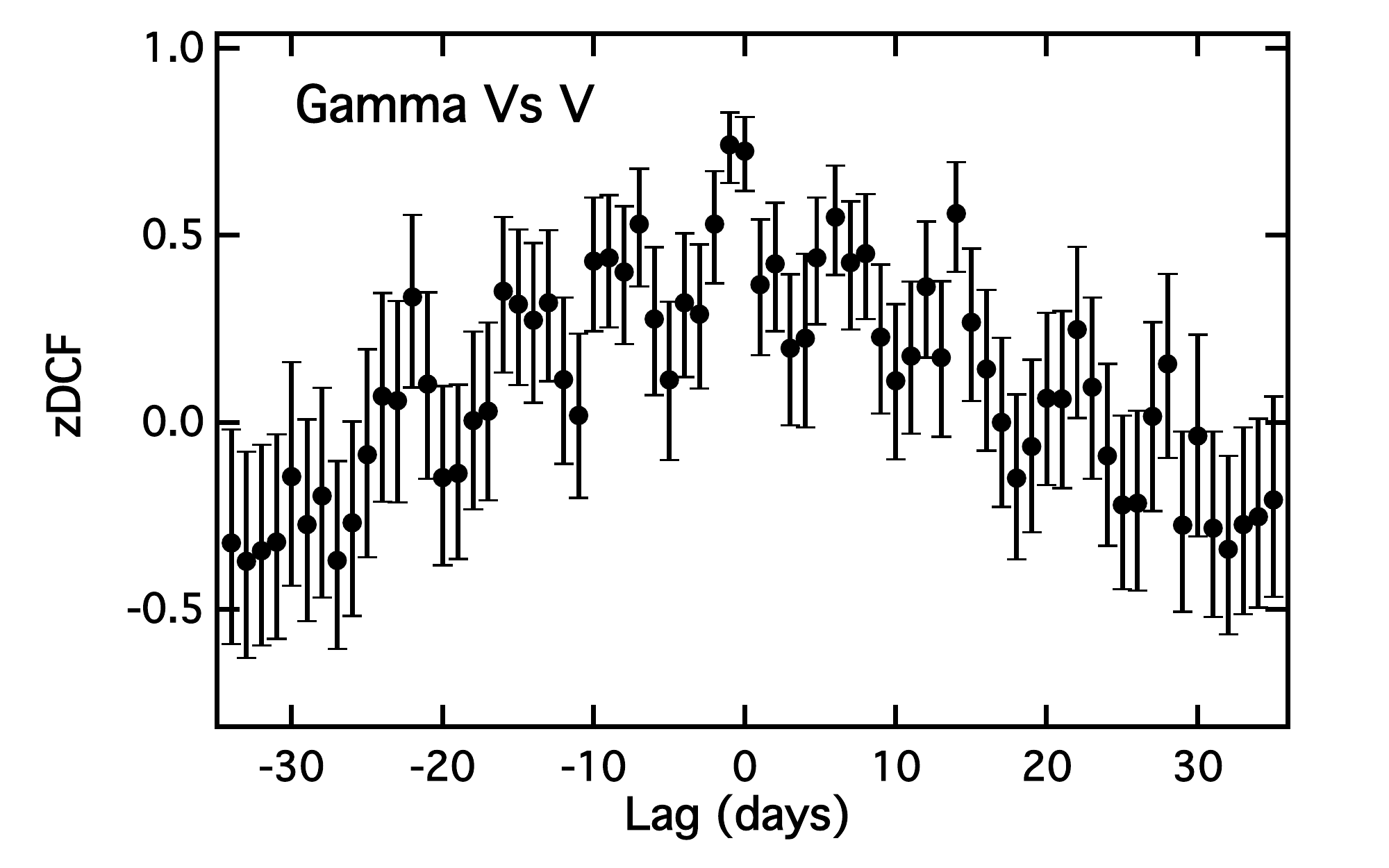}
%dcf_GX_black.pdf
\includegraphics[width=0.32\textwidth]{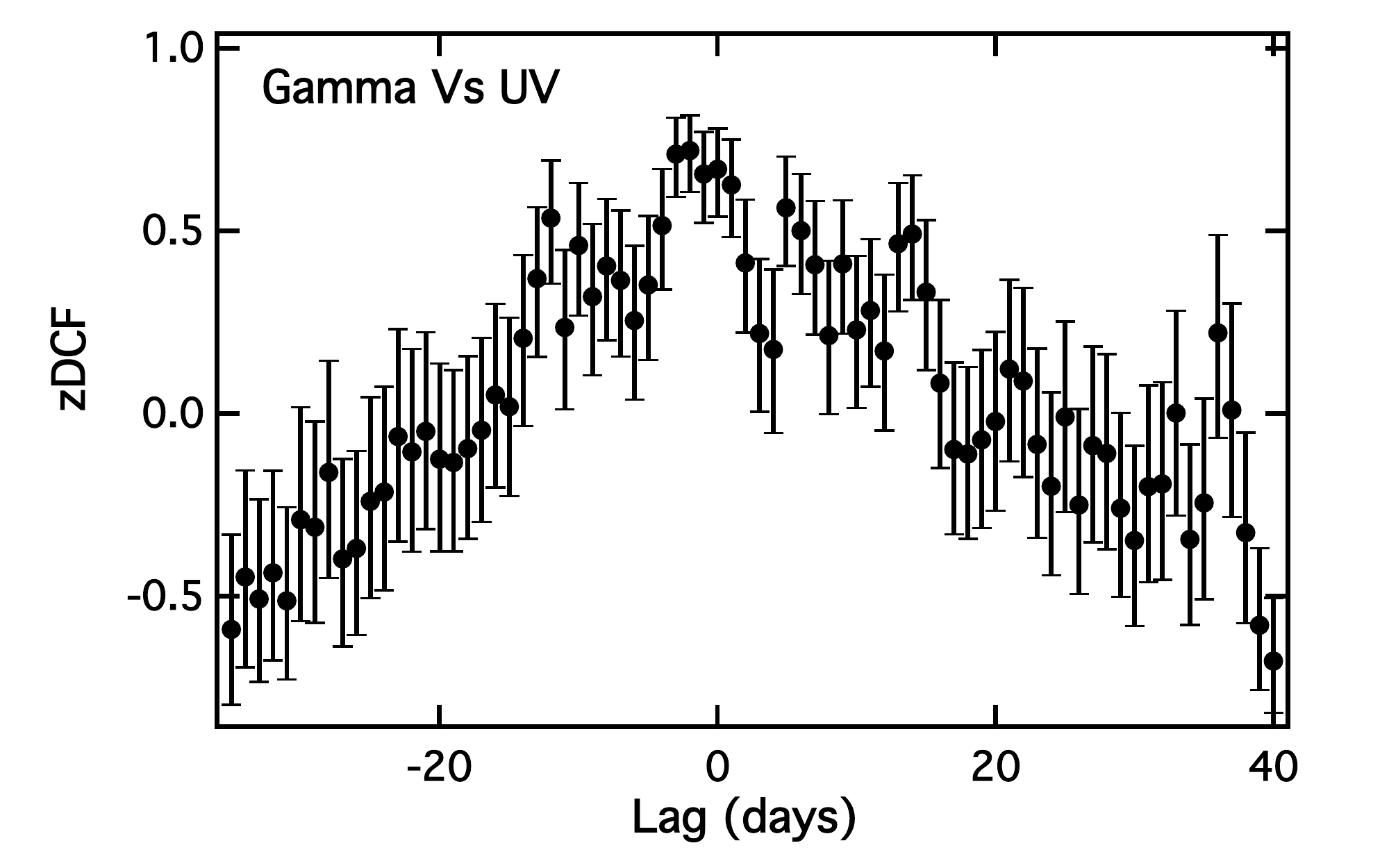}
%dcf_GUV_black.pdf
\includegraphics[width=0.32\textwidth]{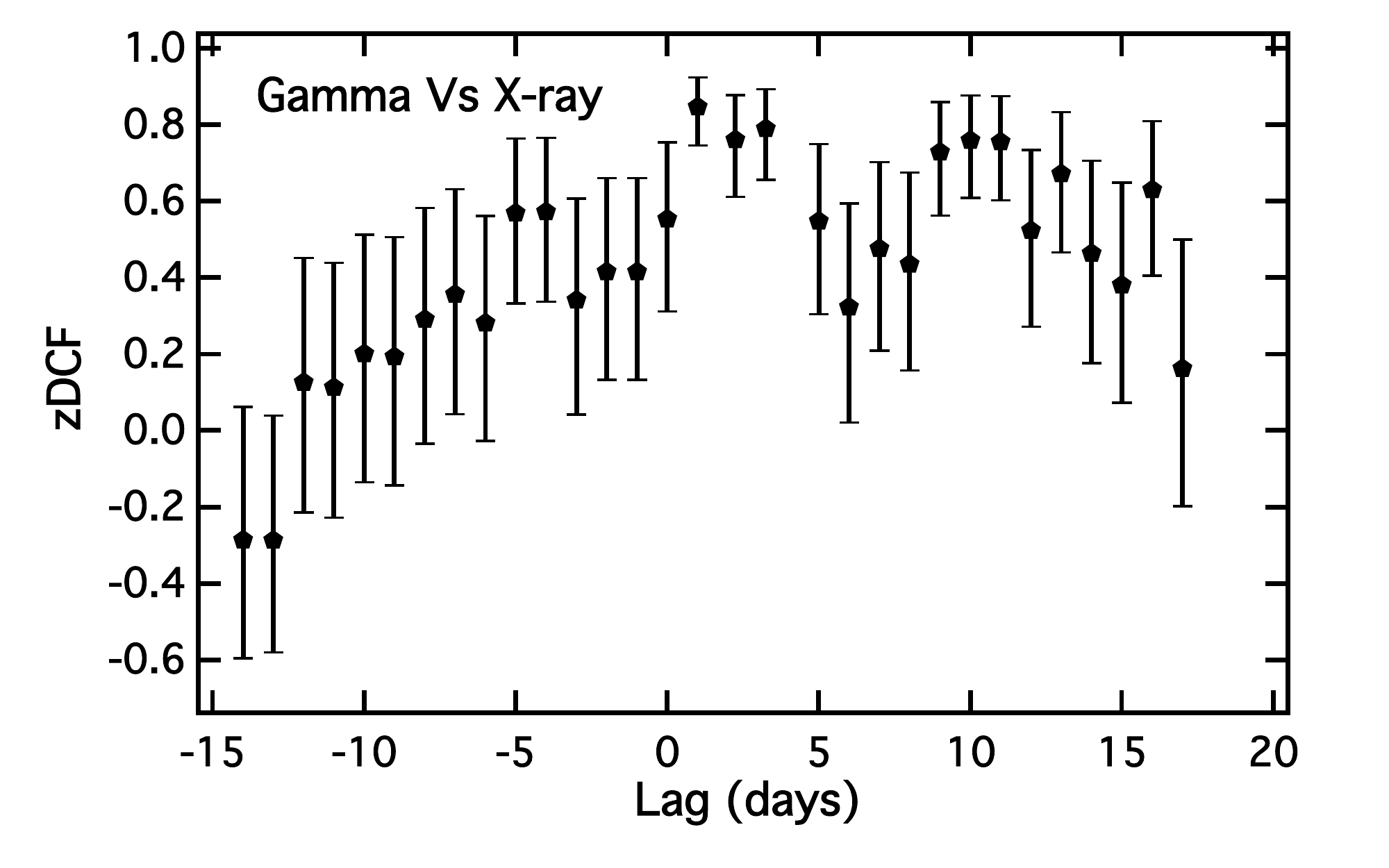}

\caption{ Discrete correlations (zDCF) for CTA\,102 between $\gamma-$rays and optical,  UV and X-ray fluxes, respectively, during 2016 November- 2017 January. 
} 
\label{dcf}
\end{figure*}

The Figure \ref{dcf} shows the correlation plots for $\gamma-$ray $\&$  optical V-band (left panel), $\gamma-$ray \& UV (center panel) and $\gamma-$ray $\&$ X-ray (right panel) light curves. As can be noticed from these zDCF plots, strong  correlations are seen in all the cases  with small or no lags (within errors/bin sizes).  

The correlation plot for $\gamma-$ray and optical flux shows a very strong correlation without any significant lag ($\gamma-$ray lags by about  1 day, which is within the errors) between the variable emissions, with DCF peak value as 0.75.  Similar behaviour is shown by $\gamma-$ray versus UV (Fig. \ref{dcf}, center panel). The correlation exhibited between  $\gamma-$ray and optical emissions during this outburst  is similar to the one reported for CTA\, 102 by \citet{Larionov2016} during 2012 outburst where they found a remarkable similarity in two energy regimes behaviour.  It was proposed that the two emission regions were co-spatial. They, however, reported 1 hr as the time lag between $\gamma-$ray and optical emissions in the post-outburst phase and claimed the high energy emission to be produced under SSC process where seed photons are provided by the synchrotron low energy emission. Albeit, during the outburst, a linear relationship between the flux variations in the two energy bands was noticed  indicating external seed photons getting up-scattered  to higher energies during the flaring. The variable flux was  explained based on the varying viewing angle, which results in change in Doppler boosting of the flux \citep{Larionov2016}. During this phase, they also noticed harder when brighter trend in both, the optical and $\gamma-$ray non-thermal emissions. During the flaring, we also notice the similar behaviour as discussed in earlier section.

It is to be noted that if the $\gamma-$ray emission is due to inverse-Compton scattering of soft photons off the same electrons which produce  the optical radiation, then its variations are likely to be simultaneous or delayed with respect to those in the optical radiation, due to light travel time delay, based upon non-thermal flares caused by the shocks moving down the  jet \citep{sokolov2004}.  Such behaviour was also noticed in FSRQ 4C38.41 \citep{Raiteri2012} and  3C 454.3\citep{bonning2009, vercellone2010}. However, opposite behaviour,  $\gamma-$ ray flux variations leading the optical ones, is also reported, e.g., in the FSRQs PKS 1510-09 by \citet{Abdo2010b} and for 3C 279 it was noticed by \citet{hayashida2012}. It could be explained based on the idea  of faster decrease in the  external seed photon energy density involved  in inverse Compton process as compared to the decay in the magnetic field energy density, which plays dominant role in synchrotron emission, along the jet.  \citet{marscher2014} discussed such a scenario when correlated emissions behave in such complex fashion based on the turbulence effects in the jet model (TEMZ). The changing magnitude and direction of a turbulent magnetic field largely leads to the variation in the synchrotron emission and hence a variable optical flux. It will not affect the high energy  gamma-ray flux. As an alternative explanation, high energy gamma-ray emitting region could be better aligned with respect to the line of sight, as compared to optical emitting region, leading to higher Doppler boosting of high energy flux.

% Figure 6.5

Such strong correlation between $\gamma-$ray and IR/optical/UV variations has been noticed for many other sources  as well \citep{Cohen2014,  bonning2009, raiteri2011, vercellone2010, jorstad2013} and their co-spatial origin was inferred. It has also been seen in many cases that nature of the correlated variations between two emitted fluxes  changes with epochs. Different values of the correlation parameters could be due to differential alignment of the one emitting region with respect  to the other at different epochs \citep{villata2007, villata2009, raiteri2011}. It could, perhaps be due to different processes and/or different particle populations being involved in the activity. In the case of FSRQ 3C454.3, \citet{bonning2009} found strong correlation between $\gamma-$ray and IR/optical emission, with very short lag and a larger fractional variability towards larger wavelengths. They concluded that emissions are co-spatial and while low energy IR/optical emissions were due to synchrotron radiation, higher energy emissions were due to change in the injection of higher energy electron population instead of ambient thermal photons (which would lead to larger $F_{var}$ at higher energies). 

The zDCF between $\gamma-$ray and X-ray flux (Figure \ref{dcf}, right panel) shows significant correlation, with DCF value as 0.85 where  the former leads the later by  about one  day. However, considering the bin size and data gaps resulting in larger errors, it is safer to say that these are correlated, with time lags within error. The two emissions, therefore originate from the same or close-by regions. Apart from this, the X-ray emission shows rather strange behaviour if one looks at the flare-to-flare behaviour of X-ray lightcurve. The rapidly rising $\gamma-$ray flare centered around MJD 57745 (2016 December 23) has counter-parts in optical and UV but X-ray flux shows decreasing trend. The slow decay of the flare was also traced well in these two bands, while the X-ray started rising and peaked, the $\gamma-$ray flux reached minimum. Even after the major $\gamma-$ray flare at MJD 57751 (2016 December 29), while all the fluxes decayed slowly, X-ray flux started increasing again, peaking when all the fluxes had reached their minima. The last major flare in $\gamma-$ray on MJD 57760 (2017 January 7) peaked with one day lag w.r.t. optical/UV bands while X-ray flux formed a plateau. Such an intriguing behavior of the X-ray flux during the high flux state seen in this source, has been noticed earlier for several other blazars \citep{Cavaliere2017, nav1es2017, carnerero2017}.

It should be noted that during 2004-05 high activity, X-ray, optical and radio emission were not correlated\citep{Osterman2009}. With no significant lags in our case, all the emission regions, therefore, appear to be at least co-spatial and can be explained by the inhomogeneous jet model in which a shock propagates down the jet, interacting with the plasma over densities or stationary cores distributed randomly in the jet, leading to the emission at progressively longer wavelengths \citep{bland1979, MarscherGear1985}. Such strong correlation between high energy emission and those at lower ones suggests that low energy emission is dominated by synchrotron emission, with minor contribution from accretion disk/host galaxy, while the higher energy emissions are due to up-scattering of the synchrotron photons with a possible contribution from external population of the seed photons, by the high energy electrons responsible for synchrotron emission\citep{jorstad2013}.
\smallskip

\noindent{\bf Average flux levels in CTA\,102 during 2016 outburst}
\smallskip

The optical outburst activity during November 2016--January 2017, in CTA\,102 is recently reported by \citet{Bachev2017} showing the brightest magnitude as R$=$11.43 mag as on December 23, 2016. The authors reported the averaged R-band magnitude during the flaring period as 12.56 mag.  
%However, our optical observations from MIRO shows that the source reaches its historical brightness, showing R$=$10.92$\pm$0.01 on December 29, 2016. 
%{\bf Nav put the values as obtained in 2016 outburst for optical, UV- I have put tentative values. Also, see if you can get values from 2012 outburst- Larionov/Bachev papers)}
It is to be noted that unprecedented flare fluxes apart, the average flux levels detected in the present study on CTA\,102  are also the highest ever reported. We report the average brightness levels for $\gamma-$ray, X-ray, UV and optical as $2.39\times10^{-6} \ ph \ cm^{-2}\ s^{-1}$,  $2.75\times10^{-11} \ erg \ cm^{-2}\ s^{-1}$, 12.45 mag (W1), 13.44 mag $\&$ 11.63 (R; MIRO), respectively. However, the brightest values achieved by the blazar CTA\,102 during 2016 outburst are unprecedented as noted here,  $1.04\times10^{-5} \ ph\ cm^{-2}\ s^{-1}$ ($\gamma-$ray),  $5.1\times10^{-11} \ erg \ cm^{-2}\ s^{-1}$ (X-ray;XRT), 12.45 mag (W1;UVOT), 13.44 mag(B;UVOT), 10.92 (R; MIRO).  The long-term consistent increase in the optical flux (from about 14.5 in 2016 June to 10.92 mag in R-band on 2016 December 29) could be due to a systematic increase in the magnetic field strength, particle density or a systematic decrease of the external radiation energy density that reduces the probability of energy losses. Alternatively, these could result from a decrease in the viewing angle, which leads to stronger Doppler boosting of the jet emission  \citep{Bachev2017, Larionov2016} at higher frequencies. 

\section{Conclusions}
\label{sec_conc}
After the 2012 huge optical flare, FSRQ CTA\,102 flared up again, beginning 2016 June 8. This outburst with several flares covering EMS, is discussed using the data from Fermi-LAT, Swift-XRT/UVOT and ground based observatories, Steward \& MIRO. The study reveals that the source was in its historically brightest phases in the $\gamma-$ray, X-ray, optical and UV bands. There were several instances when the flux increased manifolds due to strong flaring events which lead to short term (3d to 10d) and long-term (30d or longer) variability trends. The MIRO data suggest the source to show strong day-to-day variations. We noticed correlated, quasi-simultaneous multi-wavelength emissions with a delay of not more than a few days, indicating to their co-spatial origins. The larger values of fractional variability towards higher frequencies suggest more activity at higher energies. Based on the halving time scale ($\tau=$4.72 hrs) in $\gamma-$ray flux, the emission region size is estimated to be as $8.76\times 10^{15}$ cm. Taking the jet opening angle as $0.7^\circ$ \citep{Jorstad2005}, we estimate the distance of $\gamma-$ray emission region from the central super-massive black hole as $5.58\times10^{16}$ cm which lies near the boundary of BLR region.

It is proposed that the consistent flaring activity, which kept the flux levels high, is caused by the frequent injection of plasma into the jet and its interaction with quasi-stationary core. The strong shocks lead to the subsequent gain in the particle energy, on either side of the shock, leading to the enhanced flux levels. As an alternative explanation, viewing angle could be decreasing which will lead to enhanced emission throughout EMS due to increase in Doppler boosting. The base flux could be  enhanced by perturbation or as a series of discrete, possibly overlapping flares, originating at shock fronts as shocks travel down the jet.

We infer a harder when brighter spectral behavior in X- and  $\gamma-$ray emissions. A relatively softer trend during major $\gamma-$ray flares, when near peak,  suggests an efficient cooling mechanism due, perhaps, to the high particle density in the vicinity of BLR \citep{sikora2002} leading to the rapid quenching of particles before their attaining higher energies.  On the other hand, higher energy radiation/particles might be escaping the emitting region. Rapid flares could be the result of interaction of blobs with the plasma inhomogeneities/irregularities in the emission region. 

In the multiwavelength $lcs$, X-ray emission behaves rather strangely at times, which could  be due to its differential orientation with respect to LOS or different origin of the X-ray emission.

\begin{acknowledgement} 
This work is supported by the Department of Space, Government of India. We gratefully thank the Referee for fruitful suggestions. We also thank Drs. S. Ganesh, P. Grandi, E. Torresi, and M. Cappi for discussions. Data from the Steward Observatory spectro-polarimetric monitoring project,  supported by Fermi Guest Investigator grants NNX08AW56G, NNX09AU10G, NNX12AO93G, and NNX15AU81G, are used. We used  Enrico, a community-developed Python package to simplify Fermi-LAT analysis. This work made use of data supplied by the UK Swift Science Data Centre at the University of Leicester \citep{Evans2007}.
\end{acknowledgement}  

\bibliographystyle{aa} 
\bibliography{referencesUniv}

\end{document}